\documentclass[10pt,twocolumn]{article}

\usepackage{amsmath}
\usepackage{amssymb}
\usepackage{graphicx}
\usepackage{dcolumn}
\usepackage{bm}

\usepackage[superscript]{cite}

\usepackage[small]{titlesec}

\usepackage[utf8]{inputenc}
\usepackage[T1]{fontenc}
\usepackage{mathptmx}
\usepackage{abstract}
\usepackage{etoolbox}
\usepackage{bm}
\usepackage[uline]{hhtensor}
\usepackage{amsfonts}
\usepackage{xcolor}

\usepackage{hyperref}
\hypersetup{
    colorlinks=true,
    linkcolor=blue,
    filecolor=blue,
    citecolor = black,      
    urlcolor=cyan
    }

\usepackage[margin=0.8in]{geometry}

\usepackage{draftwatermark}
\SetWatermarkText{\bf PREPRINT ~~~ PREPRINT ~~~ PREPRINT   ~~~ PREPRINT}
\SetWatermarkFontSize{1cm}
\SetWatermarkAngle{90}
\SetWatermarkHorCenter{20.5cm}
\SetWatermarkColor[gray]{0.75}

\begin{document}

\title{Realizability conditions for relativistic gases with a non-zero heat flux}

\author{Stefano Boccelli\thanks{Corresponding author: stefano.boccelli@polimi.it}\\
\textit{\normalsize Department of Mechanical Engineering, University of Ottawa, ON, Canada.}\\[2ex]
James G.~McDonald\\
\textit{\normalsize Department of Mechanical Engineering, University of Ottawa, ON, Canada.}
}

\date{\textbf{Accepted for publication} in Physics of Fluids, 2022}

\twocolumn[
  \begin{@twocolumnfalse}

    \maketitle

    \begin{abstract}
    \noindent This work introduces a limitation on the minimum value that can be assumed by the energy of a relativistic gas 
in the presence of a non-zero heat flux. 
Such a limitation arises from the non-negativity of the particle distribution function, and is found by solving the Hamburger moment problem.
The resulting limitation is seen to recover the Taub inequality in the case of a zero heat flux, but is
more strict if a non-zero heat flux is considered.
These results imply that, in order for the distribution function to be non-negative, 
(i) the energy of a gas must be larger than a minimum threshold; 
(ii) the heat flux, on the other hand, has a maximum value determined by the energy and the pressure tensor;
and (iii) there exists an upper limit for the the adiabatic index $\Gamma$ of the relativistic equation of state,
and that limit decreases in the presence of a heat flux and pressure anisotropy, asymptoting to a value $\Gamma = 1$.
The latter point implies that the Synge equation of state is formally incompatible with a relativistic gas showing a heat flux, except in certain gas states.
    \end{abstract}

    \vskip2ex
  \end{@twocolumnfalse}
]

\saythanks

\section{Introduction}\label{sec:introduction}

From kinetic theory arguments, Taub\cite{taub1948relativistic} has shown that the kinetic energy density, $\rho \varepsilon$, 
and the hydrostatic pressure, $P$, of a relativistic gas must respect the inequality
\begin{equation}
  \left( \rho \varepsilon - 3 P \right) \rho \varepsilon \ge \rho c^2 \, ,
\end{equation}

\noindent where $\rho$ is the mass density of the gas in the local rest frame and $c$ is the speed of light.
This inequality has proven crucial in the formulation of equations of state (EoS). 
For instance, the Taub inequality rules out the usage of an adiabatic index, $\Gamma$, equal to the classical value of $5/3$ 
for a monatomic gas, and implies instead that, in the relativistically hot gas limit, $\Gamma$ should decrease to $4/3$
(see for instance Mignone \& McKinney\cite{mignone2007equation}).
The Synge EoS\cite{synge1957relativistic} embeds these limits naturally, but the presence of Bessel functions in its formulation,
together with the need to invert it during numerical simulations in order to find primitive variables, has triggered the 
search for alternative equations of state that are at the same time accurate and reasonably easy to solve.\cite{ryu2006equation,chattopadhyay2009effects}

As discussed in Section~\ref{sec:nomenclature-theory}, the relation between the energy and the pressure (and thus, 
ultimately, the EoS) depends on the shape of the phase-space particle distribution function, $f$. 
In general, all thermodynamic variables and macroscopic quantities can be obtained as statistical moments of $f$.
Considering that $f$ is non-negative by definition, as it represents the number of particles per unit phase-space volume, 
one may expect that not all imaginable gas states (sets of macroscopic quantities) are possible, 
if the non-negativity of $f$ is to be preserved.
For instance, states characterized by a negative density or a negative energy are impossible.
However, less trivial conditions also exist and additional non-linear combinations of moments may also result to be impossible.
Such conditions can be formalized by solving the Hamburger moment problem.\cite{hamburger1944hermitian}
In this work, we aim to do this for a single-component relativistic gas, and neglecting quantum effects.

In Section \ref{sec:nomenclature-theory}, we introduce the notation employed in this work, we define 
the moments of the particle distribution function and highlight their thermodynamic interpretation.
Then, in Section \ref{sec:1D-hamburger}, we consider the simple case of a gas made of particles that 
possess a single translational degree of freedom (particles in a spatially one-dimensional world).
For such a gas, we obtain a set of necessary conditions that the moments need to respect in order to 
be realizable by a non-negative distribution function.
The Taub inequality appears naturally as one such condition.
In Section \ref{sec:phys-realiz-implications}, we consider the implications of these conditions on 
the maximum allowable heat flux and on the admissible equations of state.
Some material supporting the 1D (one-dimensional) problem, and a comparison of the Synge EoS with the results of this work, are presented in Appendix~\ref{sec:1p-maxwell-juttner-eos}.
Finally, Section \ref{sec:3p-hamburger} extends the results to a more realistic gas with $\mathsf{N}$ translational
degrees of freedom ($\mathsf{N}=3$ being the typical case). 


\section{Notation and relativistic kinetic description}\label{sec:nomenclature-theory}

In this work, uppercase indices are used to indicate four-vectors, 
and lowercase indices indicate their spatial part.
For instance, the space-time coordinates are defined by the 
four-vector $x^A = (c t, x^a)$, where $c$ is the speed of light.
The momentum four-vector of a particle is written 
as $p^A = (p^0, p^a)$, and its contraction is $p^A p_A = m^2 c^2$,
with $m$ the particle mass.

The derivations discussed in this work simplify significantly if one considers
particles with a single momentum component, $\mathsf{N} = 1$. 
This corresponds to limiting the spatial index to $a \equiv x$, and 
the metric tensor in that case is written as $\eta^{AB} = \mathrm{diag}(1,-1)$.
In this work, we start by considering results in this simplified scenario, and then extend them to 
the typical case of $\mathsf{N} = 3$ spatial dimensions, with $\eta^{AB} = \mathrm{diag}(1,-1,-1,-1)$.

In relativistic kinetic theory, 
macroscopic variables are obtained as moments of the distribution function 
(or ``phase density''), $f(x^A, p^a)$, integrated over the $\mathsf{N}$-dimensional momentum 
space.\cite{cercignani2002relativistic}
The first two moments of $f$ are the particle four-flow and the 
energy-momentum tensor.
The former reads
\begin{equation}\label{eq:N-A-kinetic-definition}
  N^A = c \int_{-\infty}^{+\infty} p^A f \, \mathrm{d}P \, ,
\end{equation}

\noindent where $\mathrm{d}P = \mathrm{d}^\mathsf{N} p/p^0$ is the invariant 
infinitesimal momentum element.
In the typical case of particles with three momentum components, 
$\mathrm{d}P = \mathrm{d}^3 p/p^0$ and the integral in Eq.~\eqref{eq:N-A-kinetic-definition} is thus triple.
The energy-momentum tensor is obtained as 
\begin{equation}\label{eq:T-AB-kinetic-definition}
  T^{AB} = c \int_{-\infty}^{+\infty} p^A p^B f \, \mathrm{d}P \, .
\end{equation}

\noindent In this work, we employ the Eckart decomposition, such that the heat flux is associated to the energy-momentum tensor. 
Also, the quantities are considered in the Lorentz rest frame (``LRF'', indicated by a subscript $R$), co-moving with the gas.
Under these premises, the moments assume a simple thermodynamic meaning.\cite{cercignani2002relativistic}
The particle four-flow becomes
\begin{equation}\label{eq:N-A-LRF-composition}
  N^A_R = (c n, \bm{0}) \, ,
\end{equation}

\noindent with $n$ the number density of the gas ($\rho = n m$ being the mass density in the rest frame), and 
where the boldface font is used to denote the spatial component of the four-vector.
The energy-momentum tensor in the LRF is
\begin{equation}\label{eq:T-AB-LRF-composition}
  T^{AB}_R 
= \begin{bmatrix}
  T^{00} & T^{0a} \\ 
  T^{b0} & T^{ab} \\ 
  \end{bmatrix}_R
= \begin{bmatrix}
  \rho e & q^a/c \\ 
  q^a/c & P^{ab} \\ 
  \end{bmatrix} \, .
\end{equation}

\noindent In the LRF, the purely time component $T_R^{00}$ represents the energy density of the gas, $\rho e$,
that decomposes as $\rho e = \rho c^2 + \rho \varepsilon$, giving the rest energy and the kinetic energy respectively.
The purely spatial components, $T^{ab}_R$, represent the $\mathsf{N}$-dimensional pressure tensor, $P^{ab}$.
Additionally, we define the hydrostatic pressure as the (scaled) trace of the pressure tensor, $P := -{P^a}_{a}/\mathsf{N}$.
For an equilibrium gas, the pressure tensor is diagonal and reads $P^{ab} = - P \eta^{ab}$.
However, this condition is not necessary in the present work, where we retain instead its general form.
The mixed time-space components $T^{0a}_R$ represent the flux of energy, and in the LRF are equal to the 
heat flux, $q^a$, scaled by the speed of light.
The contraction of the energy-momentum tensor is $({T^A}_A)_R = \rho e - \mathsf{N} P$.

As discussed by Dreyer \& Weiss,\cite{dreyer1986classical} in the classical limit, 
the relativistic thermodynamic variables $\rho \varepsilon$, $P^{ab}$ and $q^a$
recover the traditional definitions of classical kinetic theory\cite{ferziger1973mathematical} (subscript ``$\mathrm{clas}$'') within a factor 
of $\mathcal{O}(c^{-2})$.
Symbolically denoting the classical limit as ``$c \to \infty$'', we have:
\begin{subequations}
\begin{equation}\label{eq:classical-rhoe-definition}
  \lim_{c \rightarrow \infty} \rho \varepsilon = \int \frac{m v^2}{2} f_\mathrm{clas}(v) \, \mathrm{d}^\mathsf{N} v = (\rho \varepsilon)_\mathrm{clas}\, ,
\end{equation}
\begin{equation}\label{eq:classical-Pab-definition}
  \lim_{c \rightarrow \infty} P^{ab} = \int m v^a v^b f_{\mathrm{clas}}(v) \, \mathrm{d}^\mathsf{N} v = P^{ab}_{\mathrm{clas}}\, ,
\end{equation}
\begin{equation}\label{eq:classical-q-definition}
  \lim_{c \rightarrow \infty} q^a = \int \frac{m v^2}{2} v \, f_{\mathrm{clas}}(v) \, \mathrm{d}^\mathsf{N} v = q^a_{\mathrm{clas}} \, ,
\end{equation}
\end{subequations}

\noindent where $f_{\mathrm{clas}}(v)$ is the classical velocity distribution function, and where the particle velocity $v$ appears in 
place of the peculiar velocity because we have assumed to be in the LRF. 


\section{Realizability conditions in $\mathsf{N}=1$ spatial dimensions}\label{sec:1D-hamburger}

We consider here a gas composed of particles that possess a single spatial momentum component,
$p^A = (p^0, p^x)$.
In the following, we refer to such a gas as spatially one-dimensional, or ``1D''.
While not necessarily physically realistic, this case simplifies drastically the formulation and allows to 
gain a deeper understanding of the problem.
The general ``$\mathsf{N}$D'' case is considered in Section~\ref{sec:3p-hamburger}.

The distribution function, $f$, introduced in Section~\ref{sec:introduction} represents the number of particles
per unit of phase-space volume, and is thus non-negative by definition.
As known, this non-negativity causes the density, energy and pressure to be non-negative as well.
However, non-negativity also introduces further and less obvious constraints, as not all combinations of moments
are possible.
The problem of finding the states (sets of moments) that are compatible with a non-negative
distribution function is known as the Hamburger moment problem.\cite{hamburger1944hermitian}
This problem has been frequently employed in classical gas dynamics,\cite{mcdonald2013affordable} where the states
that are compatible with a non-negative distribution function are said to be ``physically realizable''. 
Our aim here is to extend this to a relativistic gas.

First, we shall compose an array $\vec{M} = (1, p^0, p^x)$, and use it to build the matrix 
$\vec{M}^\intercal \vec{M}$.
If we compute moments of the distribution function using such a matrix as a weight, we obtain 
\begin{equation}
  \matr{Y} 
= c \int \vec{M}^\intercal \vec{M} f \, \mathrm{d}P 
= c \int 
     \begin{bmatrix} 
        1   &  p^0      &  p^x \\
        p^0 &  p^0 p^0  &  p^0 p^x \\
        p^x &  p^x p^0  &  p^x p^x \\
     \end{bmatrix} 
     f \, \mathrm{d}P \, .
\end{equation}

\noindent Most of these moments have been defined in Eqs.~\eqref{eq:N-A-LRF-composition} and \eqref{eq:T-AB-LRF-composition},
while the first entry can be written as:
\begin{equation}
  Y_{1,1} 
= c \int f \, \mathrm{d} P 
= c \int \frac{p^A p_A}{m^2 c^2} f \, \mathrm{d} P 
= \frac{{T^{A}}_A}{m^2 c^2} \, .
\end{equation}
\noindent In the LRF, these moments are 
\begin{equation}\label{eq:Y-R-1p}
  \matr{Y}_R = 
     \begin{bmatrix} 
        \frac{\rho e - P}{m^2 c^2}   &   n c      &  0 \\
        n c &  \rho e  &  q/c \\
        0   &  q/c  &  P \\
     \end{bmatrix} \, .
\end{equation}

\noindent As mentioned, in this 1D case, we have defined the hydrostatic pressure as $P = P^{xx}$.
Also, we have omitted the superscript on the heat flux, meaning $q = q^x$.

The $\vec{M}^\intercal \vec{M}$ matrix is symmetric, by construction.
Also, it is easily verified to be positive semi-definite (PSD), with eigenvalues $\lambda_{1,2}=0$, and $\lambda_3 = 1 + p^0p^0 + p^x p^x > 0$.
Since it is PSD, all its principal minors are non-negative.\cite{prussing1986principal}
Considering that $\mathrm{d} P = \mathrm{d} p/p^0$ is positive, 
one obtains the following result:
if $f$ is non-negative, then the matrix $\matr{Y}$ is PSD, and all its principal minors are non-negative.  
This translates into a set of \textit{necessary} conditions for the moments that appear in Eq.~\eqref{eq:Y-R-1p}.
A gas state (a set of moments) that does not respect such conditions cannot be represented by a non-negative distribution 
function $f$ and is thus kinetically impossible.

Considering the principal minors of $\matr{Y}_R$, obtained by removing either zero, one or two rows and columns, one obtains seven different conditions for the moments.
Among these, the following three conditions are interesting and constitute the core of this work:
\begin{equation}\label{eq:conditions-i-ii-iii}
  \begin{cases}
    \rho e - P \ge 0 \ \ \ \hfill \mathrm{(C_i)} \, , \\
    \left[\rho e - P\right] \left[ \rho e \right] - \rho^2 c^4 \ge 0 \ \ \ \hfill \mathrm{(C_{ii})} \, ,\\
    {\left[\rho e - P\right]} \left[\rho e - \dfrac{q^2}{P c^2} \right] - \rho^2 c^4 \ge 0 \ \ \ \hfill \mathrm{(C_{iii})} \, . 
  \end{cases}
\end{equation}

\noindent We refer to these conditions as C\textsubscript{i}, C\textsubscript{ii} and C\textsubscript{iii}.
The first condition states that the total energy (rest energy plus kinetic energy) must be 
larger than the pressure. 
The second condition is well known and was previously obtained by Taub\cite{taub1948relativistic} 
from different arguments.
This form of condition C\textsubscript{ii} differs from the original Taub's inequality by a factor $3$ multiplying the pressure.
This factor does not appear here since we are considering a 1D gas, but is recovered in Section~\ref{sec:3p-hamburger}.
The third condition C\textsubscript{iii} appears as a generalization of Taub's result, and includes the effect of a non-zero heat flux on realizability.
All three conditions must be satisfied for the distribution function to be non-negative.

The PSD requirement for $\matr{Y}_R$ gives four other conditions, that are however less interesting.
Indeed, two of these only require the pressure and the energy density to be non-negative quantities.
As for the two remaining conditions, one is a duplicate of C\textsubscript{i} (provided that the pressure is positive), while the other reads
\begin{equation}
  \rho e \, P - q^2/c^2 \ge 0 \, .
\end{equation}

\noindent This condition is satisfied automatically by the previous Conditions (i-iii) and does not bring any additional information.
Therefore, the only conditions that we need to analyze in this work are Conditions (i-iii).

Finally, note that relations analogous to Conditions (i-iii) for higher-order moments can be obtained by considering additional entries
in the vector $\vec{M}$.
For instance, one could consider $\vec{M}=\left(1, p^0, p^x, p^0 p^x, p^xp^x, \cdots \right)$.
However, this goes beyond the scope of the present work.

In the following section, we analyze the region of moment space where all conditions are satisfied.
The implications of these conditions are then discussed in Section~\ref{sec:phys-realiz-implications}. 


\subsection{Realizability boundary in the $(\rho \varepsilon)_\star$ -- $q_\star$ space}

\begin{figure*}[htb]
  \centering
  \includegraphics[width=\textwidth]{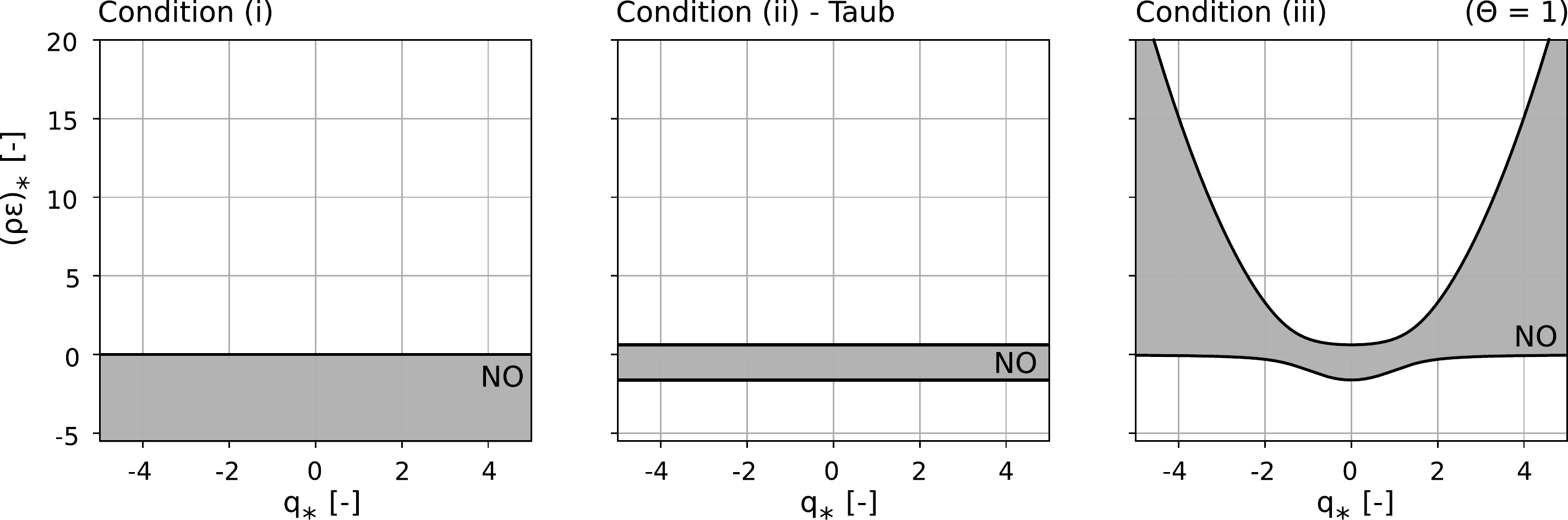}
  \caption{Conditions (i--iii) in moment space, from Eqs.~\eqref{eq:conditions-i-ii-iii-dimensionless}, for a value of $\Theta = 1$. States located in the shaded regions do not satisfy the respective conditions and therefore cannot be
           realized with a non-negative distribution function.}
  \label{fig:conditions-i-ii-iii-in-mom-space}
\end{figure*}

It is convenient to recast the conditions in Eq.~\eqref{eq:conditions-i-ii-iii} in a non-dimensional form.
One has different possibilities, such as scaling the quantities by powers of the speed of light.
However, such a constant scaling would not let any self-similarity features emerge.
Instead, we proceed as follows:
if we consider the roots of the third condition, C\textsubscript{iii}, we have
\begin{multline}\label{eq:roots-condition-iii}
  \left(\rho \varepsilon\right)_{\pm}^{\mathrm{(iii)}}
= \frac{1}{2} \left[ P + \frac{q^2}{P c^2} \right]
\\
  \pm \frac{1}{2} \sqrt{P^2 + 4 \rho^2 c^4 + \frac{q^2}{c^2} \left(\frac{q^2}{P^2 c^2} - 2 \right)} - \rho c^2 \, .
\end{multline}

\noindent The dimensionless term $(q^2/P^2 c^2 - 2)$ under the square root suggests that we employ a 
non-dimensionalization of the heat flux based on the pressure. 
In particular, if we divide Eq.~\eqref{eq:roots-condition-iii} by $P$, we obtain the following dimensionless groups (subscript ``$\star$''):
\begin{equation}\label{eq:dimensionless-groups}
  q_\star = q/(P c) \ , \ \ (\rho \varepsilon)_\star = \rho \varepsilon/P  \ , \ \  \Theta = P/(\rho c^2)  \, .
\end{equation}

\noindent As customary,\cite{chattopadhyay2009effects,mignone2005hllc}
the symbol $\Theta$ is used here to denote the pressure--density ratio, 
that expresses the non-dimensional gas temperature.
This non-dimensionalization happens to be very convenient, as it
\begin{itemize}
  \item Reduces the number of unknowns, removing the density from the picture;
  \item Provides an automatic scaling for the heat flux, allowing us to compare the results obtained 
        for different values of $\Theta$.
\end{itemize}

\noindent Conditions (i), (ii) and (iii) in dimensionless form read:
\begin{equation}\label{eq:conditions-i-ii-iii-dimensionless}
  \begin{cases}
    (\rho e)_\star - 1 \ge 0 \ \ \ \hfill \mathrm{(C_i)} \, , \\
    \left[(\rho e)_\star - 1 \right] \left[ (\rho e)_\star \right] - 1/\Theta^2 \ge 0 \ \ \ \hfill \mathrm{(C_{ii})} \, ,\\
    {\left[(\rho e)_\star - 1\right]} \left[(\rho e)_\star - q_\star^2 \right] - 1/\Theta^2 \ge 0 \ \ \ \hfill \mathrm{(C_{iii})} \, . 
  \end{cases}
\end{equation}

\noindent All three conditions of Eq.~\eqref{eq:conditions-i-ii-iii} need to be satisfied in order to have a non-negative
distribution function.
After solving for the energy, these conditions are shown separately in Fig.~\ref{fig:conditions-i-ii-iii-in-mom-space}, where we consider the special case of $\Theta = 1$.
Notice that Fig.~\ref{fig:conditions-i-ii-iii-in-mom-space} shows Conditions (i--iii) employing the kinetic energy 
$(\rho \varepsilon)_\star$ instead of the total energy $(\rho e)_\star$.

Conditions C\textsubscript{i} and C\textsubscript{ii} do not contain any information on the heat flux, 
and thus appear as straight lines in moment space.
From Fig.~\ref{fig:conditions-i-ii-iii-in-mom-space} and Eq.~\eqref{eq:conditions-i-ii-iii-dimensionless}, we notice that 
C\textsubscript{iii} recovers exactly the Taub inequality, in the special case of $q_\star = 0$.
Considering all three conditions, the most stringent one is constituted by the positive root of C\textsubscript{iii}, that reads
\begin{equation}\label{eq:positive-root-iii}
  (\rho \varepsilon)^{\mathrm{(iii)}}_{\star}= \frac{1}{2} \left[ 1 + q^2_\star + \sqrt{1 + \frac{4}{\Theta^2} + q_\star^2 \left( q_\star^2 - 2 \right)} \right] - \frac{1}{\Theta} \, .
\end{equation}

\begin{figure}[ht]
  \centering
  \includegraphics[width=\columnwidth]{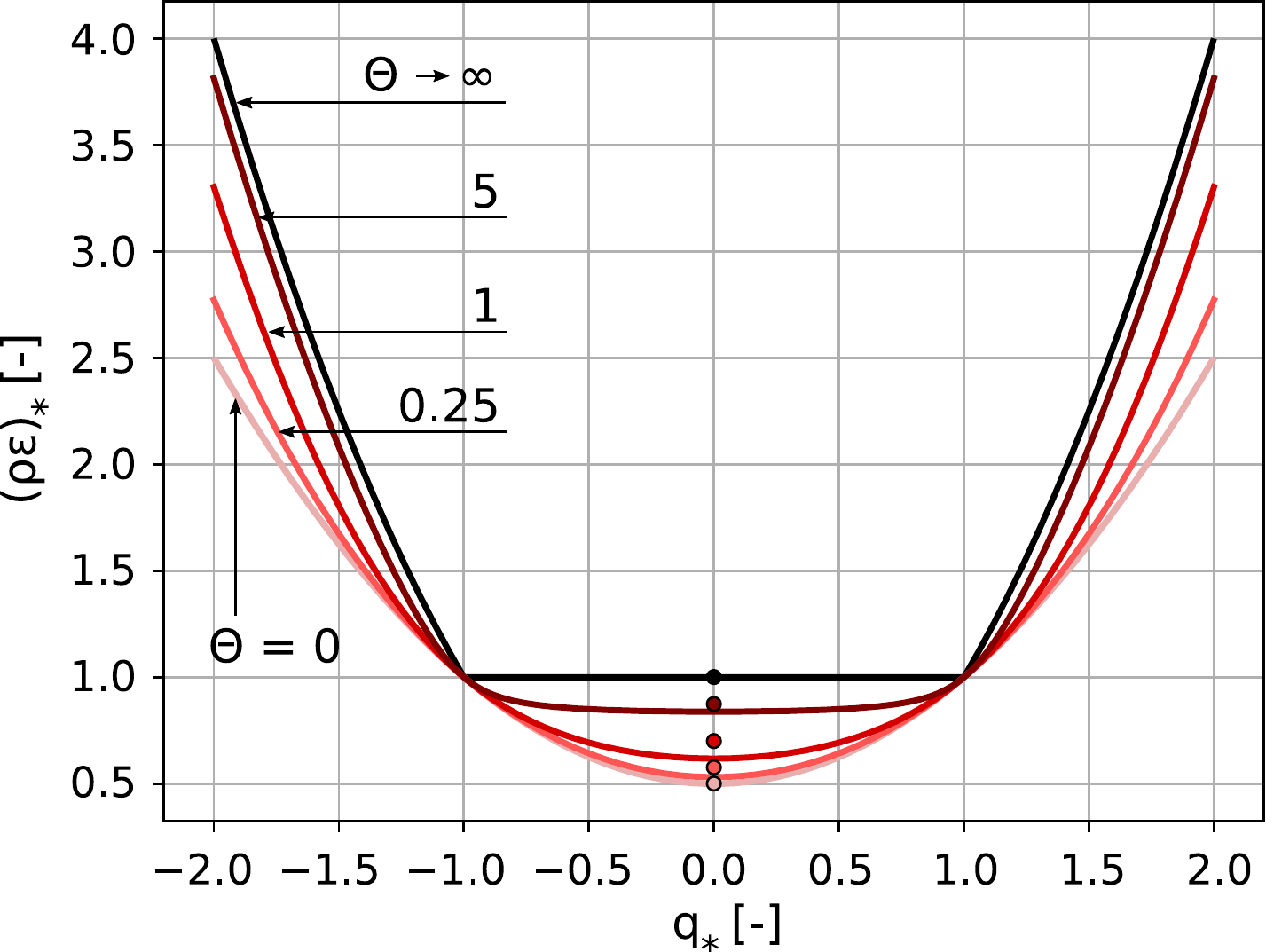}
  \caption{Realizability boundary from Eq.~\eqref{eq:condition-iii-energy} for different values of $\Theta$.
           From the bottom: $\Theta = 0, 0.25, 1, 5$ and $\Theta \rightarrow \infty$.
           Black circles at $q_\star=0$ denote the equilibrium Maxwell-J\"uttner distribution at these values of $\Theta$.}
  \label{fig:physical-realiz-boundary-equil}
\end{figure}

\noindent This condition is denoted here as the ``realizability boundary'', since only states with an energy 
\begin{equation}\label{eq:condition-iii-energy}
 (\rho \varepsilon)_{\star} \ge (\rho \varepsilon)_{\star}^{\mathrm{(iii)}}
\end{equation}

\noindent can be realized by a non-negative distribution function.
This boundary is shown in Fig.~\ref{fig:physical-realiz-boundary-equil} for various values of $\Theta$.
Thermodynamic equilibrium (the Maxwell-J\"uttner distribution) is represented by a single point on that space, that is always
above the realizability line and has a value of $q_\star = 0$. 
In Fig.~\ref{fig:physical-realiz-boundary-equil}, equilibrium is shown by black circles, for varying values of $\Theta$.
A discussion of the 1D Maxwell-J\"uttner distribution is given in Appendix \ref{sec:1p-maxwell-juttner-eos}.
All other points in the $(\rho \varepsilon)_\star$--$q_\star$ space represent different non-equilibrium states.

\subsection{Ultrarelativistic and classical limits}

For $\Theta \rightarrow \infty$ (ultrarelativistic limit), the realizability boundary is given by the line
\begin{equation}
  \lim_{\Theta \rightarrow \infty} (\rho \varepsilon)_\star^{\mathrm{(iii)}}
= \frac{1}{2}\left[ 1 + q_\star^2 + \sqrt{\left(q_\star^2 - 1 \right)^2}\right] \, ,
\end{equation}

\noindent that is a parabola truncated at its bottom, and can be rewritten as
\begin{equation}
  \lim_{\Theta \rightarrow \infty} (\rho \varepsilon)_\star^{\mathrm{(iii)}}
= \begin{cases}
    q_\star^2 \ \ \ &\mathrm{for} \ |q_\star| \ge 1 \ , \\
    1 \ \ \ &\mathrm{for} \ -1 < q_\star < 1 \ .
  \end{cases}
\end{equation}

\noindent In the case of a relativistically cold gas (``$\Theta \rightarrow 0$''), 
the realizability boundary becomes instead
\begin{equation}\label{eq:cold-gas-limit}
  \lim_{\Theta \rightarrow 0} (\rho \varepsilon)_\star^{\mathrm{(iii)}}
= \frac{1}{2} \left(q_\star^2 + 1\right) \, .
\end{equation}

\noindent Notice that, wherease this limit appears in Fig.~\ref{fig:physical-realiz-boundary-equil}, 
that does not mean that a classical gas actually does reach all such states.
Indeed, for recovering the classical behaviour, it is not sufficient to compute the limit of $\Theta \rightarrow 0$,
but one also needs to consider that $q_\star \rightarrow 0$.
This is easily understood if one considers that the natural scaling for the heat flux of a classical gas is 
$\rho v_{th}^3$, with $v_{th}$ the thermal velocity.
Therefore, as the classical regime is approached, we have 
\begin{equation}
  \lim_{\Theta \to 0} q_\star \propto \frac{\rho \, v_{th}^3}{P c} = \frac{v_{th}}{c} \to 0 \, ,
\end{equation}

\noindent where the ratio $P/\rho$ for a classical gas is proportional to the thermal velocity squared.
Therefore, the classical limit of C\textsubscript{iii} is properly obtained by setting both $\Theta \rightarrow 0$
and $q_\star \rightarrow 0$, 
\begin{equation}\label{eq:cold-gas-limit-proper}
  \lim_{\Theta \to 0, q_\star \to 0} (\rho \varepsilon)_\star^{\mathrm{(iii)}}
= \frac{1}{2} \, ,
\end{equation}

\noindent or, in dimensional form,
\begin{equation}
  \lim_{\Theta \to 0, q_\star \to 0} (\rho \varepsilon)^{\mathrm{(iii)}} = P/2 \, ,
\end{equation}

\noindent that corresponds to the only possible value of the energy for a classical gas with 
a single translational degree of freedom (adiabatic constant $\Gamma = 3$).
In other words, Condition C\textsubscript{iii} reduces to well known principles, in the classical regime.


\section{Implications of Condition (iii)}\label{sec:phys-realiz-implications}

In this section, we remark the implications of C\textsubscript{iii}.
First, as already discussed, C\textsubscript{iii} can be interpreted as a lower limit on the energy, $\rho \varepsilon$.
Second, C\textsubscript{iii} can be seen as a limit on the heat flux.
Finally, C\textsubscript{iii} can be seen to have an effect on the allowable equations of state.


\subsection{Maximum allowable heat flux}

By solving C\textsubscript{iii} for the heat flux, we obtain that a 1D relativistic gas with a given energy, $\rho \varepsilon$,
and pressure, $P$, can support at most a maximum heat flux, such that $|q| < q^{\mathrm{max}}$.
From Eq.~\eqref{eq:conditions-i-ii-iii-dimensionless}, we get
\begin{equation}\label{eq:qmax-1D1V-allowable}
  q_\star^{\mathrm{max}} = \sqrt{\frac{(\rho e)_\star \left[ (\rho e)_\star - 1\right] - 1/\Theta^2 }{ (\rho e)_\star - 1} } \, .
\end{equation}

\noindent Note that the total energy $(\rho e)_\star = (\rho \varepsilon)_\star + 1/\Theta$ 
is used in Eq.~\eqref{eq:qmax-1D1V-allowable}.
This limit can prove useful for assessing the kinetic compliance of heat flux closures and models.\cite{garcia2017heat,mendez2022dissipative,gabbana2019relativistic}


\subsection{Allowable Equations of State}

The Taub inequality is known to pose a limitation on the physically 
allowable equations of state.\cite{mignone2007equation}
C\textsubscript{iii} generalizes this by introducing the effect of the heat flux.
After introducing an index $\Gamma$ (that does not need to be constant), C\textsubscript{iii} reads
\begin{equation}\label{eq:gamma-definition-123321}
  \frac{P}{\Gamma - 1} = \rho \varepsilon \ge (\rho \varepsilon)^{\mathrm{(iii)}} = \frac{P}{\Gamma^{\mathrm{(iii)}} - 1} \, ,
\end{equation}

\noindent or
\begin{equation}
  \Gamma \le \Gamma^{\mathrm{(iii)}} = 1 + \frac{1}{(\rho \varepsilon)^{\mathrm{(iii)}}_\star} \, .
\end{equation}

\begin{figure}[htpb!]
  \centering
  \includegraphics[width=\columnwidth]{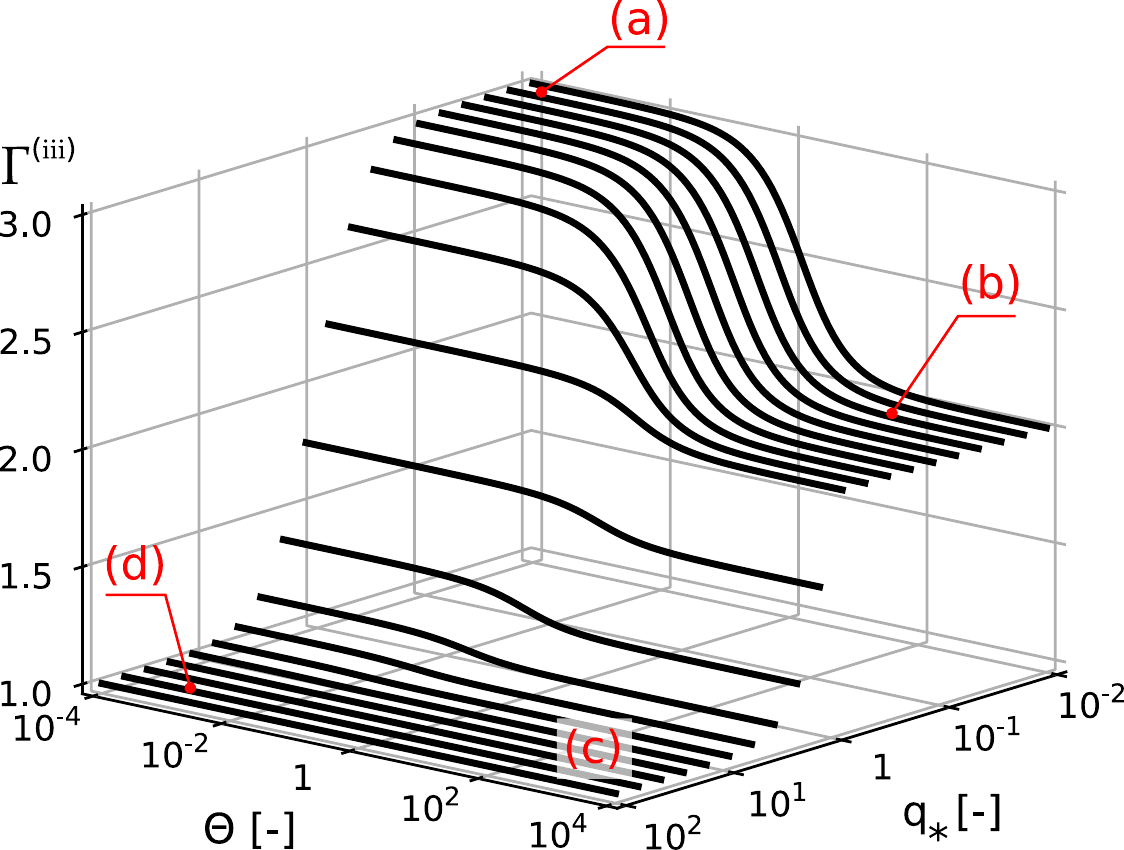}
  \caption{$\Gamma^{\mathrm{(iii)}}$: maximum physical value of the index $\Gamma$ for a 1D gas ($\mathsf{N}=1$), obtained from Condition C\textsubscript{iii}. 
           The classical limit (a) and different relativistic limits (b), (c) and (d) are identified as discussed in the text.}
  \label{fig:gamma-pstar-qstar}
\end{figure}

\noindent This gives an upper limit to the allowable values for $\Gamma$.
Note that, since the energy $(\rho \varepsilon)_\star^{\mathrm{(iii)}}$ depends on both the dimensionless temperature, $\Theta$, and the heat flux,
then $\Gamma^{\mathrm{(iii)}} = \Gamma^{\mathrm{(iii)}}(\Theta, q_\star)$.
This is shown in Fig.~\ref{fig:gamma-pstar-qstar}.
As expected, for $q_\star \rightarrow 0$, $\Gamma^{\mathrm{(iii)}}$ recovers the Taub inequality, with limits of $\Gamma = 3$ and $2$ for a cold
and hot gas respectively, as also discussed in Appendix~\ref{sec:1p-maxwell-juttner-eos}.

Condition C\textsubscript{iii} can be seen to have four asymptotic behaviors, denoted in Fig.~\ref{fig:gamma-pstar-qstar}
as regions (a),(b),(c) and (d).
In region (a), where $(\Theta,q_\star) \to (0,0)$, the classical limit is recovered. 
Moreover, the gradient of $\Gamma^{\mathrm{(iii)}}$ is zero in this region, as C\textsubscript{iii} loses all dependence on 
the heat flux. 
Region (b) is characterized by relativistic particle velocities, and by heat fluxes ranging from $q_\star \to 0$ (symmetric 
distribution function, Taub limit) to $q_\star = 1$.
In this region, the heat flux appears to play no role.
This reflects the shape of the ultrarelativistic limit of Condition C\textsubscript{iii} in moment space (see Fig.~\ref{fig:physical-realiz-boundary-equil}), 
that shows a kink at $q_\star = 1$.

As the heat flux is increased, $\Gamma^{\mathrm{(iii)}}$ asymptotes to a value of $1$.
This is expected, since large values of the heat flux, $q_\star$, can be realized only if the energy also increases
(see Fig.~\ref{fig:physical-realiz-boundary-equil}).
In the limit of $q_\star \to \infty$, and for a given and finite pressure, $P$, 
this can only be realized if the denominator of Eq.~\eqref{eq:gamma-definition-123321} goes to zero,
and thus $\Gamma \rightarrow 1$.
In Region (c), the gas is characterized by a fully relativistic distribution function with strong asymmetry.

Region (d) instead is more subtle.
In this region, the gas has a low temperature and thus may appear to be classical, yet the heat flux is relativistically significant.
Considering again Fig.~\ref{fig:physical-realiz-boundary-equil}, for a cold gas, a non-zero heat flux requires that the energy, $\rho \varepsilon$, is 
significantly super-Maxwellian.
This apparent paradox (low temperature yet high energy) could be realized for instance by distribution functions composed of two populations
of particles, such as bump-on-tail distributions.
One may think of a non-relativistic cold gas, crossed by a very dim beam of relativistic particles.
If the beam is sufficiently rarefied with respect to the bulk population, the overall temperature $\Theta = P/\rho c^2$ remains low.
Yet, the overall energy is significantly increased by the high-velocity beam. 
Moreover, the heat flux might also be significantly affected by the high-velocity particles (asymmetric distribution).
An analogous effect could be played by asymmetric and non-Maxwellian tails.

The decrease of $\Gamma^{\mathrm{(iii)}}$ implies that the Synge EoS\cite{synge1957relativistic} becomes unphysical in the presence of a significant heat flux. 
This is shown in Appendix~\ref{sec:1p-maxwell-juttner-eos}.
In the ultrarelativistic regime, the Synge EoS happens to be still valid as long as $q_\star \le 1$, but it breaks C\textsubscript{iii}
immediately after.
Instead, for small values of $\Theta$, even small values of the heat flux are such that the Synge EoS lies
above $\Gamma^{\mathrm{(iii)}}$.
This violation is not dramatic if $q_\star$ is small (for $q_\star = 0.1$, the discrepancy is below the $1\%$)
but deviations become significant soon after.

The question of what equation of state is the most suitable for such relativistic non-equilibrium cases is not trivial and ultimately
depends on the specific system of equations to be solved (e.g. traditional hydrodynamics or higher-order methods\cite{grad1949kinetic,levermore1996moment,gabbana2017kinetic,ambrucs2018high,pennisi2017relativistic,struchtrup1998projected}).
However, one may decide to employ directly the limiting condition, $\Gamma^{\mathrm{(iii)}}$, as an equation of state, for the lack of a better model.
An analogous choice was previously employed by Mignone et al.,\cite{mignone2005piecewise} in the context of an equilibrium gas with $q_\star = 0$.


\section{Realizability conditions in $\mathsf{N}$ spatial dimensions}\label{sec:3p-hamburger}

The results of the previous sections extend easily to the case of $\mathsf{N}$ spatial dimensions.
All that one has to do is to introduce additional spatial momentum components into the array $\vec{M}$.
For instance, in the case of $\mathsf{N}=3$ dimensions, one has $\vec{M} = \sqrt{c} \left(1, p^0, p^x, p^y, p^z \right)$
and the matrix $\matr{Y}_R$ reads
\begin{equation}\label{eq:Y-R-3p}
  \matr{Y}_R = 
     \begin{bmatrix} 
        \frac{({T^A}_A)_R}{m^2 c^2}   &   n c      &  0 & 0 & 0 \\
        n c   &  \rho e  &  q^x/c   & q^y/c      &  q^z/c \\
        0     &  q^x/c  &  P^{xx}   & P^{xy}     &  P^{xz} \\
        0     &  q^y/c  &  P^{yx}   & P^{yy}     &  P^{yz} \\
        0     &  q^z/c  &  P^{zx}   & P^{zy}     &  P^{zz} \\
     \end{bmatrix} \, .
\end{equation}

\noindent with $({T^A}_A)_R = ({\rho e - \mathsf{N} P})/{m^2 c^2}$.
The definitions of $\vec{M}$ and $\matr{Y}_R$ extend trivially to different values of $\mathsf{N}$:
the expressions shown in the remaining of this section are general and hold for every value of $\mathsf{N} \ge 1$.
In particular, the results of the previous section can be recovered by considering one single spatial component and setting $\mathsf{N}=1$.
The case $\mathsf{N} = 2$ can be of relevance for the study of solid-state configurations, where particles are bounded on a two-dimensional surface,\cite{novoselov2005two,watson2022relativistic} while $\mathsf{N}=3$ represents a typical gas.

As before, realizabile states make the matrix $\matr{Y}_R$ PSD.
The first two upper-left sub-matrices result in conditions that are
completely analogous to the 1D case, except that a factor $\mathsf{N}$ multiplies the pressure.
In dimensionless form:
\begin{equation}\label{eq:conditions-i-ii-iii-GENERAL-CASE}
  \begin{cases}
     (\rho e)_\star - \mathsf{N} \ge 0\ \ \ \hfill \mathrm{(C_i)} \, , \\
     \left[(\rho e)_\star - \mathsf{N} \right](\rho e)_\star - 1/\Theta^2 \ge 0  \ \ \ \hfill \mathrm{(C_{ii})} \, .
  \end{cases}
\end{equation}

\noindent As before, the second condition is the Taub inequality.
The dimensionless temperature, $\Theta$, was defined using the hydrostatic pressure, $P$, and therefore still reads $\Theta = P/\rho c^2$.
The next condition introduces the heat flux component $q^{x}$ and reads
\begin{equation}
  \left[\rho e - \mathsf{N} P \right] \left[ \rho e \, P^{xx} - \frac{q^{x\, 2}}{c^2}\right] - \rho^2 c^4 \, P^{xx} \ge 0 \, .
\end{equation}

\noindent This introduces the need to non-dimensionalize the components of the $\mathsf{N}$-dimensional pressure tensor.
We define the scaled quantities $P^{ab}_\star = P^{ab}/P$.
With this definition, we can write
\begin{equation}
  \left[(\rho e)_\star - \mathsf{N} \right] \left[ (\rho e)_\star - \frac{q^{x\, 2}_\star}{P^{xx}_\star}\right] - \frac{1}{\Theta^2} \ge 0 \, .
\end{equation}

\noindent Analogous conditions can be obtained for the remaining entries of the heat flux vector $q^a$, just by swapping the 
rows and columns of the matrix $\matr{Y}_R$.
However, we neglect them here and consider instead the full condition, obtained from non-negativity of the determinant of the full matrix $\matr{Y}_R$, that we write as
\begin{equation}\label{eq:highest-order-condition-3P-tmp1}
  \det \matr{Y}_R
= \frac{\rho e - \mathsf{N} P}{m^2c^2}
  \det
  \begin{bmatrix}
    \rho e & q^a/c \\
    q^a/c  & P^{ab}  \\
  \end{bmatrix}  
%
%
- n^2 c^2 
  \det
  \begin{bmatrix}
    P^{ab}
  \end{bmatrix}  \ge 0 \, ,
\end{equation}

\noindent where the first determinant can be computed by exploiting the block structure of that sub-matrix,
following
\begin{equation}
  \det 
  \begin{bmatrix}
    A        &  \vec{B}^\intercal\\
    \vec{B}  &  \matr{C}\\
  \end{bmatrix}
=
  \det(\matr{C}) \left( A - \vec{B}^\intercal \matr{C}^{-1}\vec{B} \right) \, ,
\end{equation}

\noindent where, using matrix notation, $A = \rho e$, $\vec{B} = \vec{q}/c = q^a/c$ and $\matr{C} = \matr{P} = P^{ab}$.
The quantity $\det(P^{ab})$ eventually cancels out being positive, and 
the highest-order condition (that we simply denote again as C\textsubscript{iii} for clarity) ultimately reads
\begin{equation}
  \left(\rho e - \mathsf{N} P \right) \left( \rho e - \frac{1}{c^2} {\vec{q}^\intercal (\matr{P})^{-1} \vec{q}} \right) - \rho^2 c^4 \ge 0  \ \ \ \hfill \mathrm{(C_{iii})}  \, ,
\end{equation}

\noindent where we have switched to matrix notation.
Notice that C\textsubscript{iii} includes the previous conditions on the heat flux components as sub-cases.
This condition is non-dimensionalized by dividing by $P^2$, giving
\begin{equation}\label{eq:condition-iii-3p-gas-dimensionless-implicit}
  \left[(\rho e)^\star - \mathsf{N} \right] \left[ (\rho e)_\star - \chi^2 \right]
- \frac{1}{\Theta^2} \ge 0 \, ,
\end{equation}

\noindent where we have introduced the following shorthand to simplify the notation:
\begin{equation}
  \chi = \sqrt{\vec{q}^\intercal_\star (\matr{P_\star})^{-1} \vec{q}_\star} \, .
\end{equation}

\noindent Solving Eq.~\eqref{eq:condition-iii-3p-gas-dimensionless-implicit} for the total energy and then computing the 
kinetic energy $\rho \varepsilon = \rho e - \rho c^2$, Condition C\textsubscript{iii} for an $\mathsf{N}$-dimensional gas reads 
\begin{equation}
  (\rho \varepsilon)^{\mathrm{(iii)}}_\star = \frac{1}{2}\left[ \mathsf{N} + \chi^2 + \sqrt{\mathsf{N}^2 + \frac{4}{\Theta^2} + \chi^2 (\chi^2 - 2 \mathsf{N})} \right] - \frac{1}{\Theta} \, .
\end{equation}

\begin{figure}[htpb!]
  \centering
  \includegraphics[width=\columnwidth]{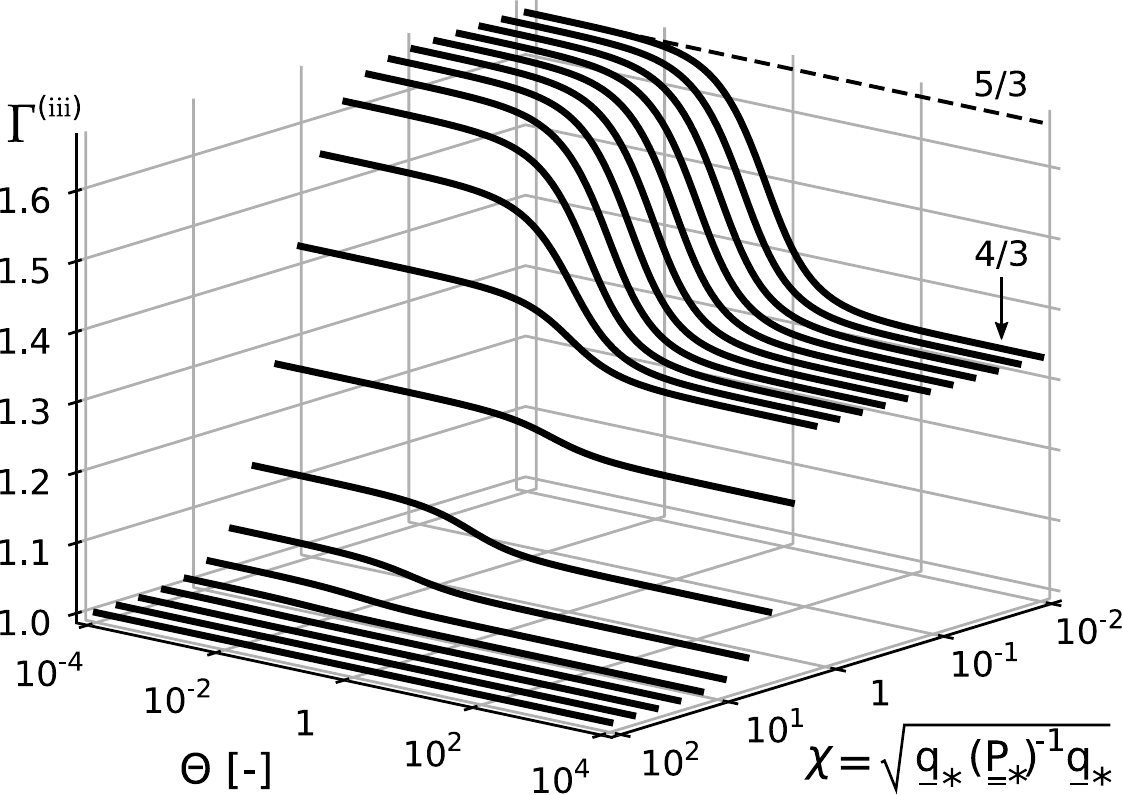}
  \caption{$\Gamma^{\mathrm{(iii)}}$: maximum physical value of the index $\Gamma$ for a 3D gas, obtained from Condition C\textsubscript{iii}.}
  \label{fig:gamma-pstar-qstar-3P}
\end{figure}

As for the 1D case, this is a lower threshold for the acceptable energies, that
translates into an \textit{upper} limit for $\Gamma$, and we have that
$\Gamma^{\mathrm{(iii)}} = 1 + 1/(\rho \varepsilon)_\star$.
The upper limit for $\mathsf{N}=3$ is shown in Fig.~\ref{fig:gamma-pstar-qstar-3P}.
For a zero heat flux, this expression recovers the known values of $5/3$ and $4/3$ for a relativistically cold and hot gas respectively.
As discussed for the 1D case, $\Gamma^{\mathrm{(iii)}}\to 1$ for relativistically large heat fluxes.

Notice that the parameter $\chi$ embeds both the heat flux and the details of the pressure tensor.
If asymmetries or anisotropies in the pressure tensor are present, these do have an effect on the realizability.
In other words, besides the heat flux, models for the gas shear stresses and/or viscosity also need to take into 
account the mentioned limits.


\section{Conclusions}

In this work, we show that the non-negativity of the particle momentum distribution function of a relativistic gas 
requires the energy to be larger than a given threshold. 
States below such a threshold cannot be realized by a non-negative distribution function.
Formally, this threshold is obtained from the solution of the 
Hamburger moment problem for a relativistic single component gas.
This realizability condition takes the form of an inequality that involves the gas pressure (through 
the dimensionless temperature, $\Theta$) as well as the heat flux and the pressure tensor anisotropies.
The Taub inequality is recovered as a special case, in the limit of a zero heat flux.
The discussed condition appears to be more strict than the Taub inequality, and 
can be interpreted as either:

\begin{itemize}
  \item A lower limit on the energy, for a prescribed value of the pressure and heat flux;
  \item An upper limit on the heat flux, if the pressure and energy are given;
  \item A limit on the realizable equations of state (EoS).
\end{itemize}

\noindent As is well known, the temperature introduces relativistic effects and limitations on the allowable equations of state:
for instance, the classically valid adiabatic index $\Gamma = 5/3$ is known to become unphysical for relativistically hot gases.
We show that the heat flux and pressure anisotropy have an analogous influence, introducing additional relativistic effects. 
This result rules out the use of the Synge equation of state whenever a sufficiently high heat flux is present.
This may also have important effects on viscous hydrodynamic numerical simulations.\cite{gabbana2020dissipative,chabanov2021general,yano2011kinetic,wang2022extensions,sahu2021magnetogasdynamic} 

It should be noted that \textit{inaccuracies} in the Synge EoS may be expected in the presence of 
a non-zero heat flux, since this would imply that the distribution function deviates from 
the equilibrium Maxwell-J\"uttner distribution, from which the Synge EoS itself is built.
However, besides possible inaccuracies, we show that this EoS is \textit{incompatible} with a positive distribution function (and thus with kinetic theory) for a range of non-equilibrium states. The actual region of realizability of this EoS is reported in the Appendix and appears to be largest in the ultra-relativistic limit. 
On the other hand, for a cold gas, even small values of the heat flux are such that the Synge EoS violates the realizability condition.


\section*{Data availability statement}
Data sharing is not applicable to this article as no new data were created or analyzed in this study.


\section*{Acknowledgments}
We wish to thank L.~P.~Quartapelle and I.~Hawke for the enlightening
conversations.
Funding for this project was provided by the Natural Sciences and Engineering Research Council of Canada through grant 
number RGPAS-2020-00122.  
The authors are very grateful for this support.


\appendix

\section{Maxwell-J\"uttner distribution for $\mathsf{N}=1$ and its equation of state}\label{sec:1p-maxwell-juttner-eos}

The Maxwell-J\"uttner (MJ) distribution function for a gas with $\mathsf{N}=1$ momentum spatial components is easily obtained 
following the derivations by Cercignani \& Kremer\cite{cercignani2002relativistic} and Dreyer,\cite{dreyer1987maximisation}
together with the following properties of the modified Bessel functions of the second type, $K_{\nu}(x)$:
\begin{equation}
  K_{\nu + 1}(x) - K_{\nu - 1}(x) = \frac{2 \nu}{x} K_\nu(x) \, .
\end{equation}

\noindent Ultimately, the 1D Maxwell-J\"uttner distribution is obtained as
\begin{equation}\label{eq:MJ-1D-asddsa}
  f_{\mathrm{MJ}}^{\mathrm{1D}} = \frac{n}{2 m c K_1(1/\Theta)} \exp\left( -\frac{\gamma(p^x_R)}{\Theta} \right) \, ,
\end{equation}

\noindent where $\gamma(p^x_R)$ is the Lorentz factor, $p^x_R$ is the spatial component of the particle momentum 
evaluated in the Lorentz rest frame, and $\Theta = P/\rho c^2$ as for the 3D case.
With respect to the typical 3D MJ distribution, in the 1D case one has Bessel functions of lower order and no factor $\Theta$ appears at the
denominator.
The equation of state can be obtained by considering the contraction of the energy-momentum 
tensor, ${T^A}_A$, and by evaluating the energy and the pressure from a direct integration of the distribution in Eq.~\eqref{eq:MJ-1D-asddsa}.
Ultimately, the 1D EoS for an ideal gas at equilibrium takes the form
\begin{equation}
\rho \varepsilon - P = \rho c^2 \left[ \frac{K_0(\zeta)}{K_1(\zeta)} - 1 \right] \, ,
\end{equation}

\noindent with $\zeta = 1/\Theta$, and thus
\begin{equation}\label{eq:gamma-eq-1D-Synge}
  \Gamma = \left[1 + \frac{\rho c^2}{P} \left(\frac{K_0(\zeta)}{K_1(\zeta)} - 1\right) \right]^{-1} + 1 \, .
\end{equation}

\begin{figure}[htpb!]
  \centering
  \includegraphics[width=\columnwidth]{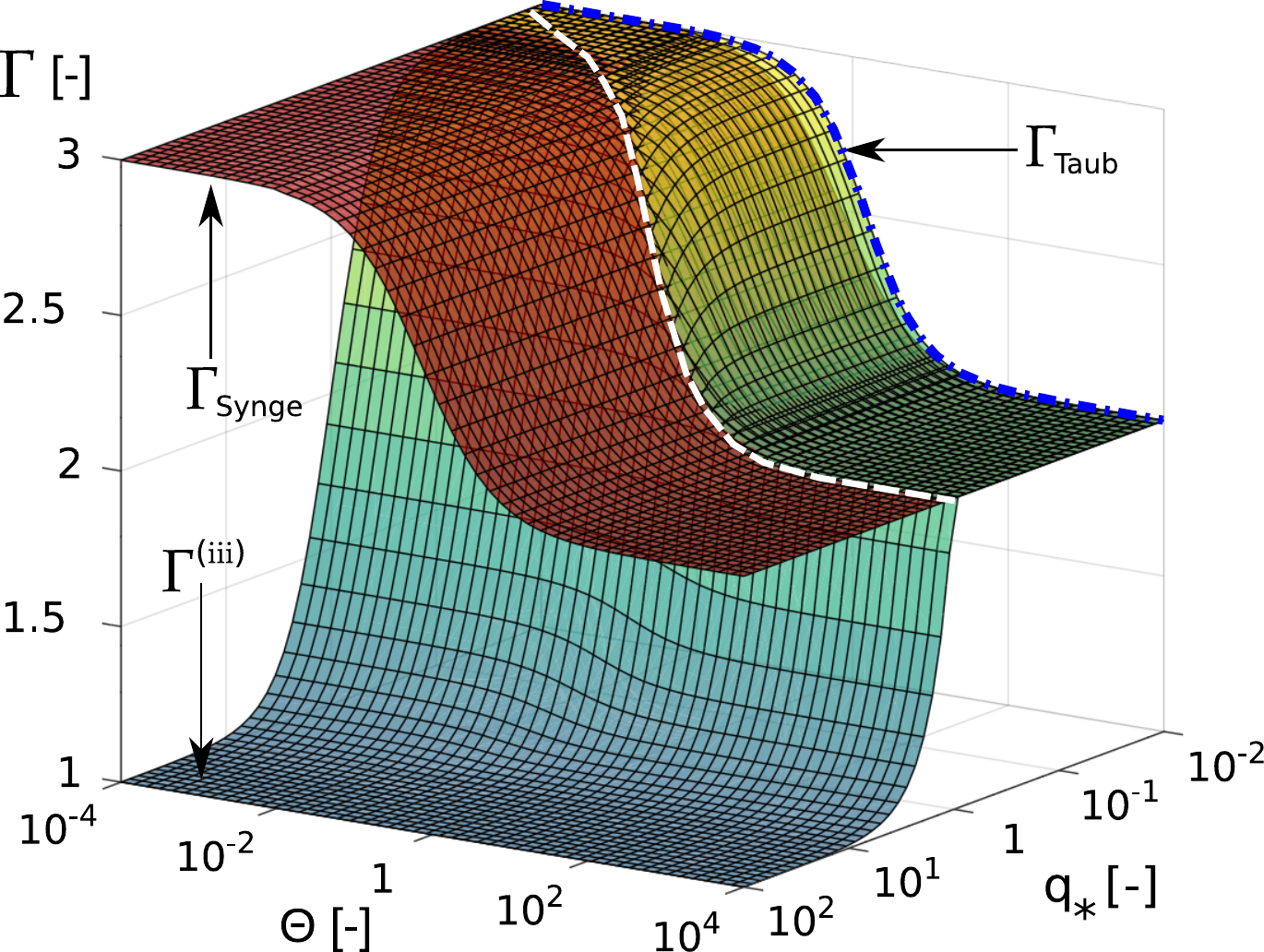}
  \caption{Synge equation of state, $\Gamma_\mathrm{Synge}$, realizability limit, $\Gamma^{\mathrm{(iii)}}$, and Taub inequality, $\Gamma_\mathrm{Taub}$, for a gas in $\mathsf{N}=1$ spatial dimensions. 
  The white dashed line delimits the region where the Synge EOS is incompatible with a positive distribution function ($\Gamma_\mathrm{Synge} > \Gamma^{\mathrm{(iii)}}$).}
  \label{fig:gamma-iii-vs-Synge-1D-gas}
\end{figure}

\noindent This equation takes the expected value of $3$ for a classical gas and $2$ in the ultrarelativistic limit.
By repeating the same derivation for a 3D gas, one obtains the typical 3D Maxwell-J\"uttner distribution function and the 
Synge equation of state,\cite{synge1957relativistic} with Bessel functions $K_2$ and $K_3$.
Figure~\ref{fig:gamma-iii-vs-Synge-1D-gas} shows the Synge EoS for a one-dimensional gas, from Eq.~\eqref{eq:gamma-eq-1D-Synge}.
In the figure, the Synge EoS is plotted in the $\Theta-q_\star$ plane, in order to compare it with 
the realizability limit, $\Gamma^{\mathrm{(iii)}}$, obtained in this work. 
Since the Synge EoS is obtained at equilibrium, it does not depend on the heat flux, and its gradient along the axis $q_\star$ is zero. 
It can be seen that the Synge EoS:
\begin{itemize}
    \item Respects the Taub inequality (Fig.~\ref{fig:gamma-iii-vs-Synge-1D-gas} for $q_\star \to 0$);
    \item Respects condition C\textsubscript{iii} only in a limited part of the $\Theta-q_\star$ plane.
\end{itemize}

%

\end{document}